\documentclass[]{pasj01}

\Received{21-Feb-2023}
\Accepted{31-Mar-2023}
 
\newcommand{\thetasun}{\theta_\odot}

\begin{document} 

\title{ 
\LETTERLABEL 
The Asymmetric Sunrise Effect on Thales' Alleged Measurement of the Sun Angular Size }

\author{Jorge \textsc{Cuadra}\altaffilmark{1,2}}
\altaffiltext{1}{Departamento de Ciencias, Facultad de Artes Liberales, Universidad Adolfo Ib\'a\~nez, \\ Av. Padre Hurtado 750, Vi\~na del Mar, Chile}
\altaffiltext{2}{N\'ucleo Milenio de Formaci\'on Planetaria - NPF, Chile}
\email{jorge.cuadra@uai.cl}


\KeyWords{History and philosophy of astronomy -- Sun: fundamental parameters}

\maketitle

\begin{abstract}
Reports from the 2nd and 3rd Century AD attribute the first measurement of the angular size of the Sun to Thales of Miletus, in the 6th Century BC.  
Cleomedes, also in the 2nd Century AD, described a method to perform the measurement, based on timing the duration of the sunrise.
Several modern authors have suggested Thales used Cleomedes' method, but others are skeptical of the connection.
Here I present an objection that has not been discussed in the literature, namely, that the proportionality between the size of the Sun and the duration of sunrise is not constant, but changes with latitude and time of the year, due to what I call the ``asymmetric sunrise effect''.
I show that this effect is large enough to have prevented Thales from obtaining the roughly accurate recorded value.
\end{abstract}


\section{Introduction}

The Sun is one of the few astronomical objects that can be resolved by the naked eye.  However, measuring its small angular diameter\footnote{There is a $<2\%$ annual variation due to the eccentricity of the Earth's orbit, which we can neglect for the purpose of this study.},
 $\thetasun=0.53^\circ$, was technically very challenging in Antiquity.  The earliest existent account of this measurement is a report by Archimedes in the third century BC, who blocked the light of the Sun with cylinders, and even took into account that the human pupil has a finite area. He managed to constrain the Sun's angular size to be $0.45^\circ < \thetasun < 0.55^\circ$ \citep{Shapiro}.

A clever alternative to the direct geometrical measurement is to time the sunrise, i.e.\ measure the duration of the time $\Delta t$ it takes for the Sun to appear completely above the horizon, as this quantity will be proportional to the Sun's angular diameter.  The procedure can be carried out without modern technology, for instance by using clepsydras, since what is required is to compare $\Delta t$ with a standard interval, such as the duration of a whole day and night cycle.  

The measurement and its result were mentioned by Cleomedes in the second century AD:

``{\it For when the Sun's size is measured out by means of water clocks, it is determined as 1/750th part of its own circle: that is, if, say, 1 kuathos\footnote{1 kuathos is about 40 ml \citep{BowenTodd}} of water flows out in the time it takes the Sun to rise completely above the horizon, then the water expelled in the whole daytime and nighttime is determined as 750 kuathoi.}''  \citep{BowenTodd} 

If we consider the Sun's ``own circle'' to be a great circle on the celestial sphere, Cleomedes gives a size of $360^\circ / 750 = 0.48^\circ$, only a 10\% smaller than the actual value.

There are two other reports from this epoch, attributing the first measurement of the angular size of the Sun to Thales of Miletus, who had lived in the sixth century BC, three centuries before Archimedes.  The first one, from Apuleius (second century AD) mentions that Thales had  ``{\it devised a marvellous calculation [...] showing how often the sun measures by its own size the circle which it describes}''.
The second, from Diogenes Laertius (third century AD) even gives the numerical result that ``{\it the size of the sun [is] one seven hundred and twentieth part of the solar circle}''  \citep[who gives both quotations]{Grady}, meaning he would have measured $\thetasun=0.5^\circ$.

Many modern authors have linked these reports, and mention that Thales may have used Cleomedes' method, or simply state that he did \citep{Tannery, Hultsch, Panchenko, Maza, Arianrhod}.
However, most experts on this subject are at best skeptical of the connection.  Some of these authors conclude that Thales is unlikely to have obtained the size at all \citep{Dicks, Graham, Grady}, while others propose that he may have done so with a geometrical measurement \citep{Wasserstein56, Rossetti}.
Their arguments are varied, but none is insurmountable.

\citet{Heath} and \citet{Graham} argue that there are no reports about Thales' measurement in the seven centuries before Apuleius, and that Archimedes in the third century BC claimed his contemporary Aristarchus was the first one to measure the size of the Sun.  Against this argument, \citet{Rossetti} reminds us that earlier texts were not readily available at those times.  He writes  ``{\it it [is] reasonable to suppose that Apuleius might have had access to information other than that which Aristarchus and later Archimedes could rely on}''.  Also, according to \citet{Panchenko}, the translation of Archimedes' text is debatable, ``{\it Archimedes does not speak about a `discovery' [...] His words imply only that Aristarchus cited this ratio on his own authority}''.

According to \citet{Heath} and \citet{Couprie}, in Thales' cosmology the Earth was flat and covered by a hemispherical firmament, so in his view the Sun could not pass under the Earth in order to make a full circle. 
\citet{Grady}, however, disagrees with these authors and attributes to Thales the idea of a spherical firmament and a circular path for the Sun.
Alternatively, \citet{Couprie} prefers to ``{\it  ascribe this discovery [...] to Anaximander because it was he who taught that the orbit of the sun [...] passes under the earth as well, making a full circle}''.  Since Anaximander was Thales' follower and successor, which one of the two would have made the measurement is not as important as whether any of them did. 
The same argument can be given to \citet{Chaikovskii}, who writes that Thales ``{\it lacked the concepts of a circular orbit and uniform motion in it}'', since Anaximander is likely to have had them.

The third objection refers to the supposedly poor accuracy of the timing method to determine the duration of the sunrise.    \citet{Couprie} and \citet{Grady} both describe procedures in which a small vessel is filled with water during the sunrise duration, and then a pair of vessels have to be alternately filled, every couple of minutes, with the same quantity of water throughout a full day and night cycle.  \citet{Grady} rightly states that ``{\it it is a cumbersome method}'',  ``{\it most unlikely [to] have provided an accurate result}''.  However, in the appendix I describe a simpler procedure Thales could have used, which we could expect to work at the $\approx10\%$ accuracy level of the known ancient figures.   
Earlier claims such as that of \citet{Neugebauer},   ``{\it it would be necessary, e.g., to guarantee an accuracy of 1/1000 in the measurement of the daily outflow}'' seem to be a misunderstanding based on the idea of measuring the day and night cycle with a precision better than the sunrise duration, 
while in reality the ratio and not the difference between those two quantities is what needs to be measured.\footnote{Perhaps the confusion is due to the debate on whether Presocratic astronomers could have determined the equinoxes by comparing the day time duration of consecutive days \citep{Dicks}.}
\citet{Neugebauer} refers also to Ptolemy's rejection of all timing methods for astronomical measurements \citep{Wasserstein, Hannah}.  However, Ptolemy did require a much higher accuracy, as he was interested in determining whether the apparent sizes of the Sun and the Moon are equal to each other, and measuring their change with time \citep{Toomer}.  Again, Ptolemy was interested in what we could call a differential measurement, while Thales would have been aiming for a relative one.

To summarise, the objections to Thales having used Cleomedes' method to measure the size of the Sun are not fundamental, and some of them have been already contested in the literature.  The remaining issue of the poor accuracy is due to assuming a very cumbersome experimental setup and to unnecessarily stringent requirements.  

In this letter I want to point out a new, astronomy-based objection to Thales alleged measurement, namely, that the duration of the sunrise is not fixed, but depends both on the location of the observer and the time of the year.  Therefore, the constant of proportionality between $\thetasun$ and $\Delta t$ is not fixed.  As far as I know, this effect has not been discussed rigorously in the context of Cleomedes' method, so in the remainder of this article I will show its magnitude and argue that Thales would not have been able to correct for it.


\section{Duration of sunrise}

In order to calculate the duration of a sunrise, we need to consider the apparent rotation of the sky, and project it on the vertical direction, considering both the inclination of the celestial sphere relative to the horizon, and the position of the rising Sun.
We follow \citet[p 50,]{Smart} and state the rate at which a celestial body rises at the horizon as
\begin{equation} 
\frac{dz}{dt} = \frac{360 ^\circ}{24 h} \sin A \cos \phi = \frac{15"}{1s} \sin A \cos \phi,
\label{eq:dzdt_Aphi}
\end{equation}
where $z$ is the altitude, $t$ is the time, $A$ the azimuth angle (measured from the North), and $\phi$ the latitude of the observer.  Since the angular factors are always smaller than unity, the fastest rate is achieved when $A=90^\circ$ and $\phi=0$, i.e.\ when the Sun is rising exactly due East, which happens only at the equinoxes, and when the observer is at the Equator. 

The azimuth angle at the horizon can be calculated from the cosine formula of spherical trigonometry as \citep[p 47]{Smart} 
\begin{eqnarray} 
\cos A = \sin \delta \sec \phi, \\  \textrm{(for $z=0$)} \nonumber
\end{eqnarray}
where $\delta$ is the declination of the Sun.  Putting both equations together, we obtain
\begin{equation} 
\frac{dz}{dt}\bigg\vert_{z=0} = \frac{15"}{1s}  \cos \phi \sqrt{1-\sin^2 \delta \sec^2 \phi}.
\label{eq:dzdt}
\end{equation}

 \begin{figure}
    \centering
    \includegraphics[angle=0,width=0.45\textwidth]{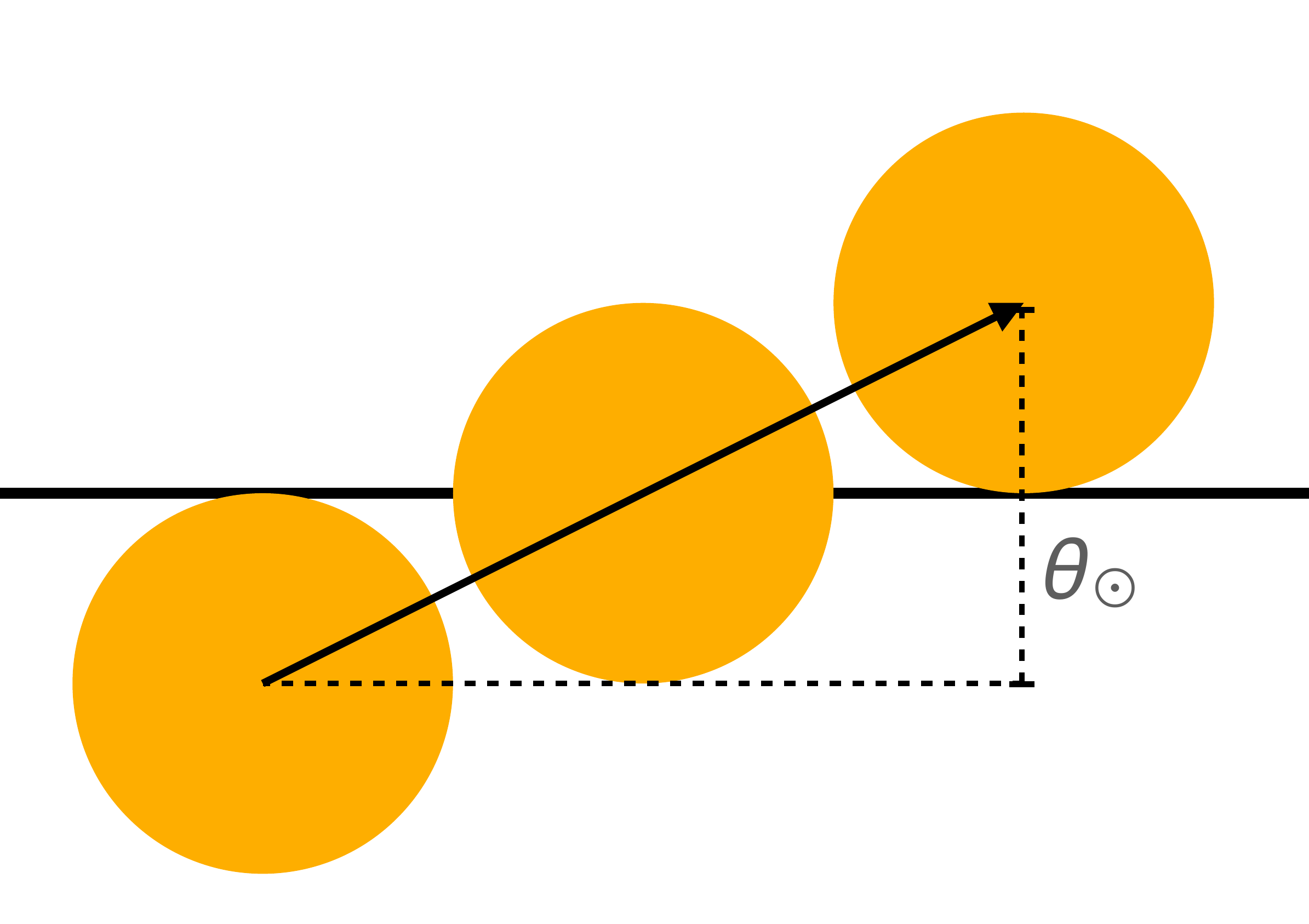}
    \caption{Schematic illustration of a sunrise, from the moment the Sun first touches the horizon, until it is completely above it.  During this time the Sun moves a distance $\thetasun$ in the vertical direction.}
    \label{fig:sunrise}
\end{figure}

The sunrise interval is defined by the Sun diameter transversing vertically the horizon, i.e.\ by a $\Delta z = \thetasun$, as shown in Fig.~\ref{fig:sunrise}.  Given the small angular size of the Sun, we can use the differential in eq.~\ref{eq:dzdt} as a fraction and calculate the duration of sunrise as
\begin{eqnarray} 
\Delta t & = \frac{1s}{15"} \frac{1}{\cos\phi} \frac{1}{\sqrt{1-\sin^2 \delta \sec^2 \phi}} \thetasun  \nonumber \\ & =127 s \frac{1}{\cos\phi} \frac{1}{\sqrt{1-\sin^2 \delta \sec^2 \phi}} .
\label{eq:deltat}
\end{eqnarray}
 It can be shown that the minimum sunrise duration is 127 s when $\phi = \delta = 0$, i.e.\ at the Equator during an equinox.   I call the deviation from this minimum duration the ``asymmetric sunrise effect'', since it is due to the Sun being not exactly in the East, and/or rising at an angle to the horizon\footnote{\label{ghorub}There does not seem to be a standard name for this angle.  \citet{Lyra} dub its equivalent at sunset the `ghorub angle', while R.\ Lachaume (priv.\ comm.) suggests to call its complement `occasive' or `ortive incidence angle', for the sunset and sunrise, respectively.}
  that is not $90^\circ$.   
 
\begin{figure}
    \centering
    \includegraphics[angle=0,width=0.45\textwidth]{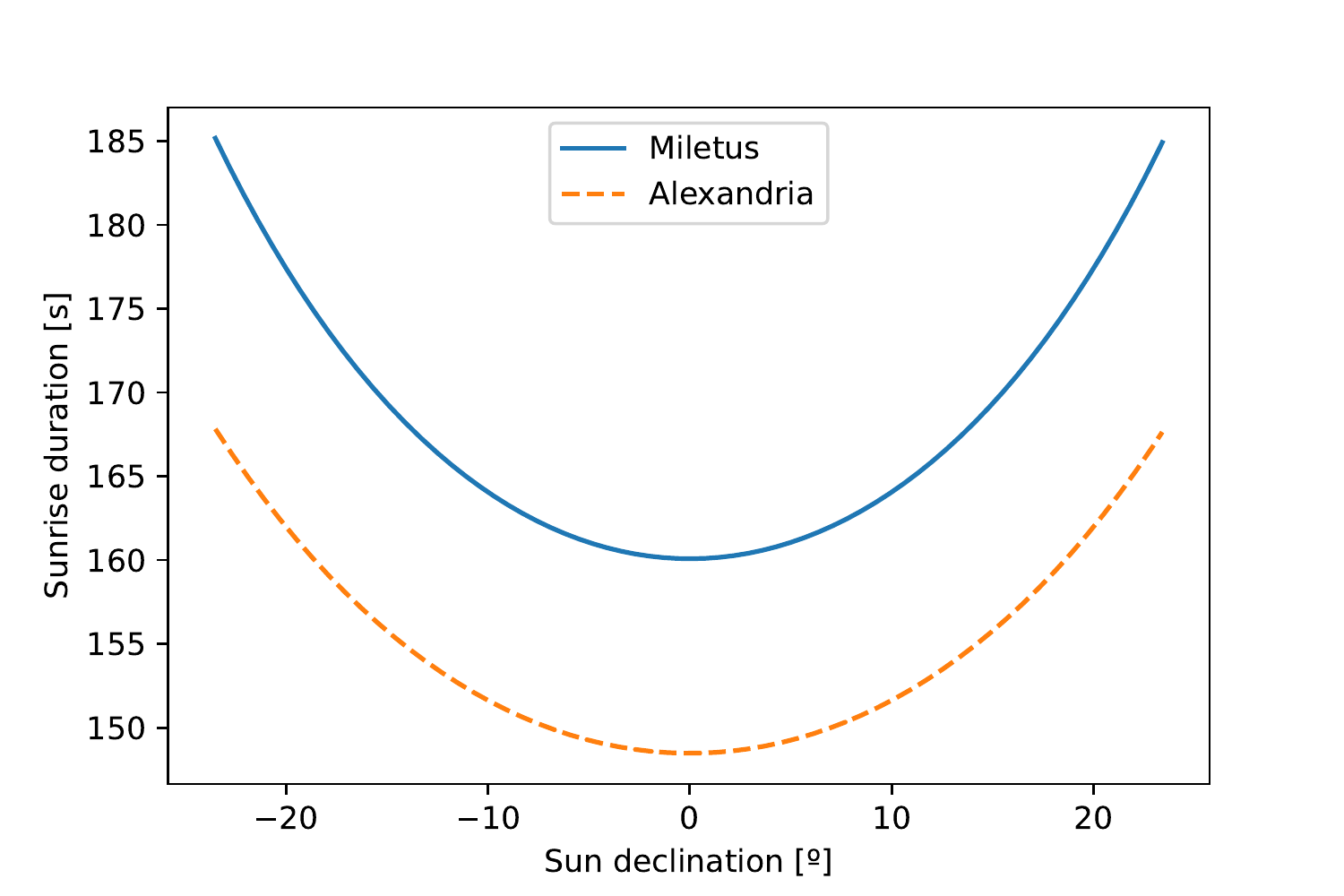}
    \caption{Duration of the sunrise $\Delta t$ in seconds, measured at the latitudes of Miletus, $\phi = 37.5^\circ$, and Alexandria, $\phi = 31.2^\circ$, for different declinations of the Sun.}
    \label{fig:deltat_Miletus}
\end{figure}

 Let us evaluate this function for the latitudes where Thales and Cleomedes were likely located, Miletus, $\phi = 37.5^\circ$, and Alexandria, $\phi = 31.2^\circ$.  Figure~\ref{fig:deltat_Miletus} shows the duration of the sunrise $\Delta t$, measured from those locations, as a function of the Sun declination $\delta$.  The latter quantity changes through the year from $0$ at the equinoxes to $\pm \varepsilon = \pm 23.5^\circ$ (the Earth's obliquity) at the solstices.  As a result, the duration of sunrise goes from 160 s to 185 s in Miletus, and from 148 s to 168 s in Alexandria.

\section{What would have Thales measured}

The timing method, as described by Cleomedes, assumes implicitly that the Sun describes a great circle on the celestial sphere, and that it rises perpendicular to the horizon.  	If that were so, one could calculate
 \begin{equation}
 \thetasun^\perp  = \frac{\Delta t}{24 h} 360^\circ =  \frac{\Delta t}{1s} 15",
 \end{equation}
where the $^\perp$ superscript is a reminder that this is a result based on a wrong assumption. 

\begin{figure}
    \centering
    \includegraphics[angle=0,width=0.45\textwidth]{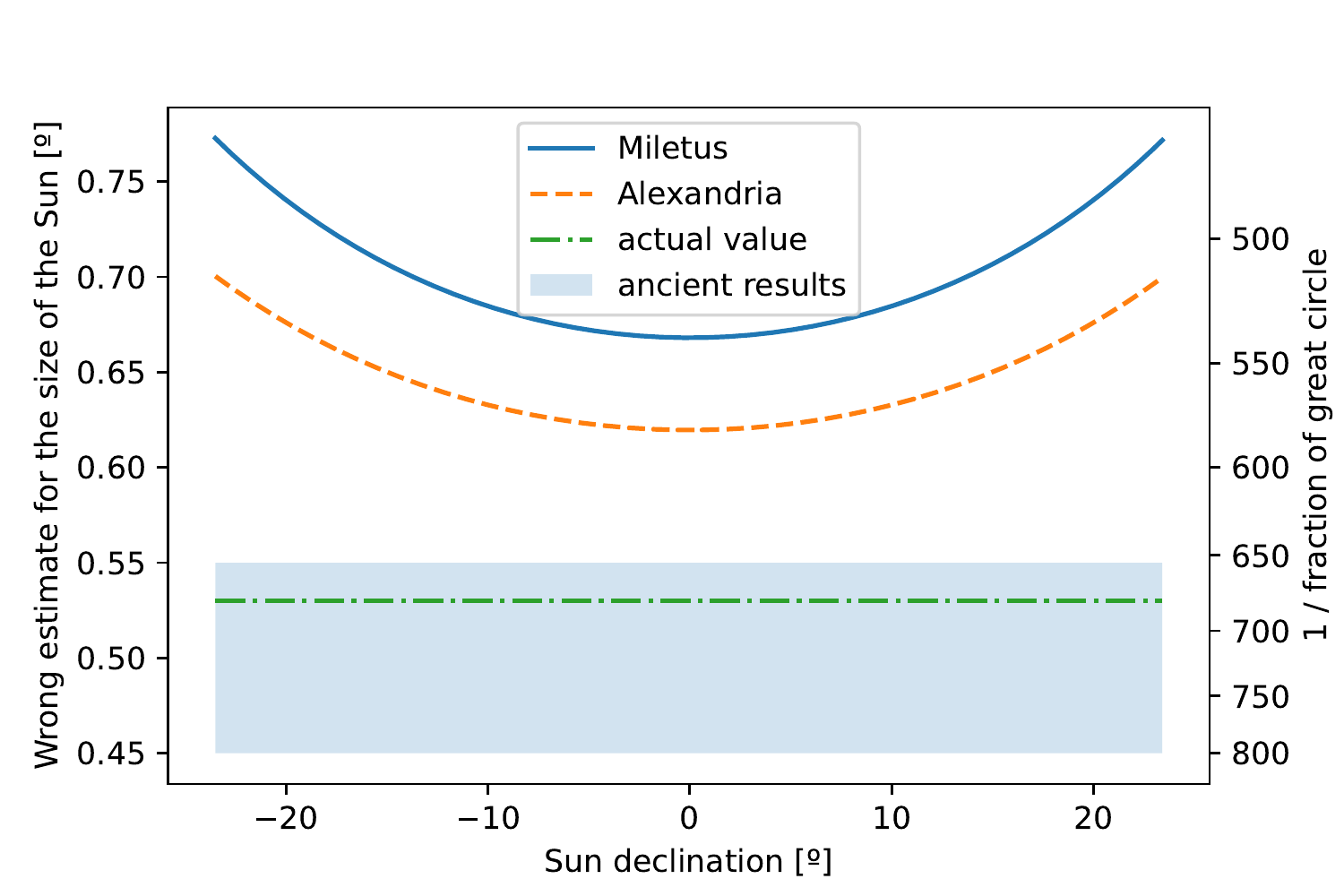}
    \caption{Wrong estimate of the angular size of the Sun  $\thetasun^\perp$  following Cleomedes' method, measured at the latitudes of Miletus, $\phi = 37.5^\circ$, and Alexandria, $\phi = 31.2^\circ$, for different declinations of the Sun.  The vertical axes show the result in degrees and in parts of a great circle.  For comparison, the actual value of the Sun as well as the range of ancient results quoted in the introduction are also shown.}
    \label{fig:thetaperp_Miletus}
\end{figure}

If an observer measures $\Delta t$ and tries to estimate the angular size of the Sun using the latter equation, they will obtain, based on eq.~\ref{eq:deltat},
 \begin{equation}
 \thetasun^\perp  = \thetasun \frac{1}{\cos\phi} \frac{1}{\sqrt{1-\sin^2 \delta \sec^2 \phi}}.
 \end{equation}
So, as expected, only at the Equator during an equinox the observer would obtain the right angular size for the Sun.  In general, they would overestimate this quantity.  I show in Fig.~\ref{fig:thetaperp_Miletus} that the values measured would be in the ranges $ 0.67^\circ - 0.77^\circ$ and $0.62^\circ - 0.70^\circ$, for Miletus and Alexandria respectively, depending on the time of the year. 
  Assuming Thales performed such measurement in his home town, he would then have overestimated the size of the Sun by between 26 and 46\%.\footnote{The overestimate is never large enough from Greek latitudes to account for the angular size of the Sun required in the interpretation of Anaximander's numbers as distances to the celestial wheels \citep[Ch.~9]{Couprie}, nor for Aristarchus puzzling value of $2^\circ$ \citep{Heath}.  To actually reach that value, $\phi>60^\circ$ is required. It is hard to imagine people at those latitudes implicitly assuming a vertical sunrise, as the ghorub angle (see footnote~\ref{ghorub}) is always $<30^\circ$ there \citep{Lyra}.}

\section{Discussion}
\label{sec:disc}

In this section I will comment on  previous studies that allude to the ``asymmetric sunrise effect'' and I will maintain that Thales could not have addressed its consequences.

\subsection{Previous remarks in the literature}

I have only been able to find two mentions of this effect in the literature, but neither presents it thoroughly.  
\citet[p 11, 224]{Delambre}, when discussing Cleomedes' text, says
 ``{\it They found the diameter of the Sun to be 28'48" at the time of the equinox, [...] we do not know if they were able to calculate it taking into account the angle that the equator makes with the horizon}'', and then he multiplies Cleomedes' diameter by $\cos (31^\circ$), assuming that he neglected to do so.  
 \citet{Delambre} then was aware of the $\cos \phi$ factor.  However, he omits the $\sin A$ factor, which is appropriate under his assumption that the method was used at the time of the equinox, but not in general.
 Elsewhere, \citet{Chaikovskii} only says that ``{\it the exact value of 1/720 cannot be obtained in this way if one lives north of the tropic}''.  In reality, only at the Equator the correct value can be obtained.

\subsection{Could the ``asymmetrical sunrise effect'' have been corrected?}

I would now like to consider whether it would have been possible for Thales to have used the timing method correctly, i.e.\ to apply the geometrical factors to go from a measurement of the sunrise duration to a correct angular size for the Sun.  Clearly he could not have used something like eq.~\ref{eq:deltat}, as that requires spherical trigonometry, which would not be developed  until the first century AD by Menelaus in Alexandria \citep{Neugebauer}.  He might perhaps have used a geometrical equivalent to eq.~\ref{eq:dzdt_Aphi} during an equinox, so $\sin A = 1$, and the only correction needed is the projection $\cos \phi$ due to the inclination of the celestial equator, which is the same as in plane geometry.  But even though the calculation may seem trivial, it was likely beyond the capabilities of any Greek philosopher of the 6th Century BC,  even Thales himself \citep[\S 4.1.3]{Rossetti}.
Moreover, to perform it would have required Thales to estimate the latitude of his location, even if the concept of latitude makes little sense in a flat Earth cosmology \citep{Dicks}.

In contrast, at the time of Cleomedes spherical trigonometry already existed, so these corrections could have been applied by himself or other astronomers. I agree then with the conclusion of \citet{Graham},  that the `Egyptians' Cleomedes attributes the method to were likely Alexandrian astronomers that used Menelaus results.  Finally, if he knew about the correction, why didn't Cleomedes include it in his text?  Perhaps we should not read it as a technical description of the experiment, but simply as part of his argument against Epicurus' claim of the Sun being a foot long, which would then imply that the firmament is only 750 feet around \citep{Wasserstein}.

\section{Conclusion}

I have shown that the ``asymmetrical sunrise effect'' results in an overestimate of $26-46\%$ for the angular size of the Sun, if measured by timing the duration of the sunrise from Miletus.  Therefore, Thales cannot have used this method if he indeed arrived at the roughly correct value of $\thetasun=0.5^\circ$, as recorded by Diogenes Laertus.  Either he used a different method, or the value was corrected before written down in the 3rd century AD.  While this conclusion is not new, I showed  that the ``asymmetrical sunrise effect'' is a more fundamental objection to Thales' alleged measurement than those previously raised in the literature.

\begin{ack}
I thank Wladimir Lyra and R\'egis Lachaume for an illuminating discussion on the angles of sunset, 
Patricia Ar\'evalo for her help to understand propagation of uncertainty, 
and  librarians Jos\'e Francisco Pe\~na and Susana Far\'ias for finding some of the bibliography.
I acknowledge partial support from ANID, -- Millennium Science Initiative Program -- NCN19\_171.
\end{ack}

\appendix 

\section*{A possible experimental setup and its accuracy}

Cleomedes did not write any details about the measurement procedure, only that the passage of time can be recorded by the quantity of water that outflowed of a vessel.  Based on the known technology available at the time, modern scholars have reconstructed possible ways to carry out the measurement.  As mentioned in the introduction, \citet{Couprie} and \citet{Grady} imagined a couple of vessels being filled up repeatedly during a full day and night cycle.  Counting the number of times the vessels were filled (several hundred) gives the answer to how many times the Sun fits on his own path.  This would indeed be a cumbersome method, prone to errors and likely to result in poor accuracy.

There is however no need to actually fill up and change the vessels every couple of minutes.  The comparison can be made not by {\it how many} vessels were filled, but rather by {\it how much} water flowed in the interval being measured.  Therefore, one can imagine a much simpler setup, in which water flows continuously from a source, and is collected in a small vessel during the duration of sunrise. Immediately after sunrise is completed, water from the same source is collected in a large vessel until the end of the following day's sunrise.  The ratio between the sizes of the Sun and its daily circular path will correspond to the ratio between the quantities of water collected in both vessels.  The latter can be measured afterwards, for instance by finding a cup that holds exactly the water collected in the small vessel, and counting how many of those cups are needed to hold all the water that was collected in the large vessel.  So this method would still have required filling up a cup several hundred times, but it avoids the stress of having to do it in real time, waiting each time for the cup to reach the desired level, and then immediately filling up another one.

A key ingredient in either setup is to have a continuous and constant supply of water.  Any large water container with a small hole would have provided a continuous source, but not necessarily constant, because the outflow rate decreases as the container gets empty.
If we assume a rectangular container, the outflow rate is proportional to its velocity, $v = \sqrt(gh)$, where $h$ is the height of the water and $g$ the gravitational acceleration. 
This effect was qualitatively known and partially corrected for in Egypt several centuries before Thales' time, as shown by the truncated cone design of the Karnak water-clock, built in the 14th Century BC \citep{Clagett}.  The shape of the container makes the outflow rate more even.  Similar devices have been found in Greece, dated to the 4th Century BC \citep{ArmstrongCamp}.  These were rather unique infrastructure, with sizes of meters and enough volume so they could work overnight without supervision.  We should therefore not assume that Thales, in the 6th Century BC, had access to such a large and accurate water-clock, but I would argue that he would have known that an even flow of water requires an even level in the large container.  Therefore, I will assume that Thales would have aimed to keep the large container at an approximately constant level for the 24-hour duration of the experiment.   

I will now show that, with this experimental setup, Thales could have kept the measurement uncertainty low enough to fit the ancient records.  
The first source of uncertainty is due to the uneven water flow filling the vessels.  Let's assume Thales kept the water height constant with an uncertainty of 10\%.  
The square root dependence implies a 5\% accuracy for  the water outflow rate.  Let's further assume that Thales topped up the large container 100 times during the experiment (i.e. almost every 15 minutes).  Assuming the errors are not correlated, the amount of water accumulated during the 24-hour cycle has an uncertainty of only 0.5\%.  If instead the errors were correlated, the flow would tend to be more even, decreasing the uncertainty.
Another source of uncertainty is related to the final measurement of the amount of water accumulated during the 24-hour experiment.  If Thales filled the measuring cup approximately 500 times to scoop out the water, each time with a 10\% accuracy, and again assuming his errors were uncorrelated, the number of cups can be determined with a 0.4\% uncertainty.  For this factor, if the errors were correlated, e.g., Thales tended to always not fill completely the measuring cup, the total error would be at most proportional to the error in each step, which in this case we assumed to be 10\%.

The remaining source of uncertainty is timing the sunrise itself.  \citet{Wasserstein} and  \citet{Grady} mention the refraction of the Sun as a source of error.  The effect of refraction at the horizon is to shift upwards the image of the Sun, making the sunrise happen earlier.   While there is substantial variation of this effect for different locations, seasons, and even for sunrise vs sunset of the same day \citep{SchaeferLiller, Sampson}, the few-minute duration of a sunrise is too short for the integrated conditions of the atmosphere from the observer to the horizon to change, and therefore the apparent size of the Sun is not affected \citep{Sigismondi11}.   \citet{Grady} also mentions that it is ``{\it difficult to observe the precise moment when the entire disc is in view}".  \citet{Sigismondi11, Sigismondi12} indeed describes the ``black drop effect", which creates an apparent connection between the lower part of the Sun and the horizon, which is due to the optics of the eye rather than to the atmosphere.  \citet{Sigismondi11, Sigismondi12} reports that his naked eye measurements overestimate the sunset duration in the range 1--5\%.   Finally, there is the uncertainty related to measuring the water accumulated during the sunrise.  I would assume Thales was even more careful for this part of the procedure than during the whole experiment, and so the water collection was accurate to well below 10\%.  

In conclusion, given the technology available in the 6th Century BC, measuring the angular diameter of the Sun with the 10\% uncertainty implied by the ancient records was feasible by Cleomedes' method of timing the sunrise.


\end{document}